\documentclass[twocolumn,showpacs,prb]{revtex4}
\usepackage{graphicx}%
\usepackage{dcolumn}
\usepackage{amsmath}
\usepackage{latexsym}

\begin{document}

\title{Seiberg-Witten map and Galilean symmetry violation in a non-commutative planar system}

\author{Biswajit Chakraborty} 
\altaffiliation{biswajit@bose.res.in}
\author{Sunandan Gangopadhyay}
\altaffiliation{sunandan@bose.res.in}
\affiliation{Satyendra Nath Bose National Centre for Basic Sciences, \\
Block-JD, Sector-III, Salt Lake, Kolkata - 700098, India;}
\author{Anirban Saha}
\affiliation{Department of Physics, Presidency College,\\86/1 College Street, Kolkata-700073, India.}

\date{\today}

\begin{abstract}
\noindent
An effective $U(1)$ gauge invariant theory is constructed for a non-commutative Schr\"{o}dinger field coupled to a background $U(1)_{\star}$ gauge field in $2+1$-dimensions using first order Seiberg-Witten map. We show that this effective theory can be cast in the form of usual Schr\"{o}dinger action with interaction
terms of noncommutative origin provided the gauge field is of ``background'' type with constant magnetic field. The Galilean symmetry is investigated and a violation is found in the boost sector. We also consider the problem of Hall conductivity in this framework. 
\end{abstract}
\maketitle

\section{Introduction}

The idea of noncommutative (NC) spacetime was introduced by Snyder \cite{sn} way back in 1947, though it was not pursued seriously by other workers till recently when this NC feature emerged as a consequence of studies in String Theory \cite{sz}. Issues related to the violation of Lorentz symmetry in NC systems have become important and studies have been done using NC variables or with their
 equivalent commutative counterpart obtained by Seiberg-Witten (SW) map
 \cite{sw}.

In this paper, we deal with a nonrelativistic system coupled to a $U(1)_{\star}$ gauge field on a NC plane. To avoid any nonunitarity or higher order time derivative terms in the action, we assume the absence of space-time noncommutativity
$(\theta^{0i}=0)$. This condition, though it spoils manifest Lorentz symmetry, is Galilean invariant. It is therefore interesting to look for any possible violation in Galilean symmetry of our system. We shall study the Galilean symmetry through an effective theory obtained by first order SW map \cite{sw}. 
Since first and second quantised formalisms are equivalent as far as Galilean invariant models are concerned, one can also carry out quantum mechanical analysis in first quantised formalism from the Schr\"{o}dinger equation derived from the effective theory. 
Note that the NC $\hat \psi$ field in Schr\"{o}dinger equation can have an interpretation of probability amplitude, but it is not clear that this feature will persist with the SW field $\psi$ when an effective commutative theory is obtained from the original NC theory. We argue here that unless the gauge field configuration provides a constant magnetic field, the probabilistic interpretation will not go through. This indicates the requirement of a ``background" type gauge field.

Finally, using the results of our effective theory, we observe that there is no
effect of the NC parameter $\theta$ on Hall conductivity.

\section{U(1)$_{\star}$ gauge invariant Schr\"{o}dinger action}

We consider NC Schr\"{o}dinger field $\hat \psi$ coupled with $U(1)_{\star}$ background gauge field $\hat A_{\mu}(x)$ in the noncommutative plane, the corresponding $U(1)_{\star}$ gauge invariant action (involving NC covariant derivative
$\hat D_{\mu}\star = \partial_{\mu}-i\hat A_{\mu}\star$) is
\begin{eqnarray}
\hat S = \int d^{3}x \hat \psi^{\dag}\star(i\hat D_{0} + 
\frac{1}{2m}\hat D_{i}\star \hat D_{i})\star \hat \psi ,
\label{2}
\end{eqnarray}
where the variables $\hat \psi$ (assumed to be Schwartzian \cite{sz}) compose through the star product defined as
\begin{eqnarray}
\left(\hat f\star \hat g\right)(x) = e^{\frac{i}{2}\theta^{\alpha\beta}\partial_{\alpha}\partial^{'}_{\beta}} 
  \hat f(x)\hat g(x^{'})\big{|}_{x^{'}=x}.
\label{3}
\end{eqnarray}
Under $ \star $ composition the Moyal bracket between the coordinates is 
$\left[\hat x^{\mu},\hat x^{\nu}\right]_{\star} = i\theta^{\mu\nu}$.
The equation of motion for the fundamental field $\hat \psi(x)$  is
\begin{eqnarray}
(i \hat D_{0} + \frac{1}{2m} \hat D_{i} \star \hat D_{i}) \star \hat \psi = 0 .
\label{16}
\end{eqnarray}
The $\star$--gauge invariant matter (probability) current density $\hat j_{\mu}$ following from (\ref{16}) is given by
\begin{eqnarray}
\hat j_{0} &=& \hat \rho = \hat \psi^{\dag} \star \hat \psi\nonumber\\
\hat j_{i} &=& \frac{1}{2mi} \left[\hat \psi^{\dag} \star \left(\hat D_{i} \star \hat \psi \right) - \left(\hat D_{i} \star \hat \psi \right)^{\dag}\star \hat \psi  \right]
\label{18}
\end{eqnarray}
 which satisfy the usual continuity equation $\partial_{t}\hat j_{0} + \partial_{i}\hat j_{i} = 0$; ($i=1,2$).

\section{Effective Theory in Commutative space}

In this section we construct an effective action starting from (\ref{2}) by 
using the SW map in the lowest order in $\theta^{\mu\nu}$ \cite{sw}; 
\begin{eqnarray}
\hat \psi = \psi - \frac{1}{2}\theta^{mj}A_{m}\partial_{j}\psi \nonumber\\
\hat A_{i} = A_{i} - \frac{1}{2}\theta^{mj}A_{m}\left(\partial_{j}A_{i} + F_{ji}\right)
\label{41}
\end{eqnarray}
Taking $\theta^{0i} = 0$, we substitute the above form of $\hat \psi$ and $\hat A_{\mu}$ in the action (\ref{2}) to find the $U(1)$ gauge invariant effective action, which when written in a hermitian form reads, 
\begin{eqnarray}
\hat{S} =&& \int d^{3}x \left[\left( 1 - \frac{\theta B}{2}\right)(\frac{i}{2}\psi^{\dag}\stackrel{\leftrightarrow}{D}_{0}{\psi}) - \frac{1}{2m}\left( 1 + \frac{\theta B}{2}\right)\right.\nonumber\\
&&\left.\times(D_{i}{\psi})^{\dag}(D_{i}{\psi}) + \frac{i}{4}\theta^{mj}({\psi^{\dag}}\stackrel{\leftrightarrow}{D_{j}}{\psi})F_{m0}\right.\nonumber \\  &&+\left.\frac{1}{8m} \theta^{mj}\left({\psi^{\dag}}\stackrel{\leftrightarrow}{D_{j}}{\psi}\right) \partial_{i} F_{mi} + ...\right]
\label{43A}
\end{eqnarray}
where the dots indicate missing terms involving $\partial_{\mu}F_{\nu\lambda} $, which are not written down explicitly, as they play no role in the simplectic structure of the theory.  
Since this action is not in the canonical form, the field $\psi$ in second quantised formalism does not have a canonical structure for the equal time commutation relation between $\psi$ and $\psi^{\dag}$ as $\left[\psi(x), \psi^{\dag}(y)\right] = \left(1 + \frac{\theta B}{2}\right) \delta^{2}(x - y)
$.
 This non-standard form of the commutation relation indicates that $\psi$ cannot represent the basic field variable or the wave function in the corresponding first
quantised formalism.
To identify the basic field variable, let us scale $\psi$ as 
\begin{eqnarray}
\psi \mapsto \tilde{\psi} = \sqrt{1 - \frac{\theta B}{2}}\psi
\label{4304}
\end{eqnarray}
so that the commutation relation can be cast as 
\begin{eqnarray}
\left[\tilde \psi(x), \tilde \psi^{\dag}(y)\right] =  \delta^{2}(x - y)
\label{4305}
\end{eqnarray}
and $\tilde \psi$ and $\tilde \psi^{\dag}$ can now be interpreted as annihilation and creation operators in second quantised formalism.
 So it becomes clear that it is $\tilde \psi$, rather than $ \psi$, which corresponds to the
  basic field variable in the action. It is therefore desirable to re-express the action (\ref{43A}) in terms of $\tilde\psi$ and ensure that  it is in the standard form in the first pair of terms. Clearly this can be done only for a constant $B$- field. Such a constant magnetic field can only arise from an appropriate background gauge field. 
In rest of the paper, we shall therefore consider a constant background for field strength tensor $F_{\mu\nu}$.
In this case, the above action (\ref{43A}) can be cast in the form, 
\begin{eqnarray}
\hat{S} =&& \int d^{3}x \left[(\frac{i}{2}\tilde\psi^{\dag}\stackrel{\leftrightarrow}{D}_{0}{\tilde\psi}) - \frac{1}{2\tilde m}(D_{i}{\tilde\psi})^{\dag}(D_{i}{\tilde\psi})\right.\nonumber\\&& + \left.\frac{i}{4}\theta^{mj}({\tilde\psi^{\dag}}\stackrel{\leftrightarrow}{D_{j}}{\tilde\psi})F_{m0}\right]
\label{4307}
\end{eqnarray}
where, $\tilde m = (1 - \theta B)m$ and $\tilde \psi$ 
 can now be regarded as renormalised mass and wave function respectively.
 We would like to mention that the expression for $\tilde m$ indicates that the external magnetic field $B$ has a critical value  $B_{c}= \frac{1}{\theta}$ as was observed also in \cite{sw}.
Incidentally, this relation for $\tilde m$ was also obtained earlier in the literature \cite{horvs}. The equation of motion for the fundamental field $\tilde{\psi}$ (from the action(\ref{4307})) is 
\begin{eqnarray}
(iD_{0} + \frac{1}{2\tilde{m}} D_{i}D_{i} + \frac{i}{2}\theta^{mj}F_{m0}D_{j})
\tilde{\psi} \equiv K \tilde \psi = 0 
\label{48}
\end{eqnarray}
Now substituting (\ref{41}) in (\ref{18}), we obtain,
\begin{eqnarray}
\hat j_{0} &=& {\psi}^{\dag}{\psi} + \frac{i}{2}\theta^{mj} \left(D_{m}{\psi}\right)^{\dag}\left(D_{j}{\psi}\right)\nonumber\\
\hat j_{i} &=& \frac{1}{2\tilde{m}i}\left[\left\{{\psi}^{\dag}\left(D_{i}{\psi}\right) - {\mathrm{c.c}}
\right\}\right.
\nonumber\\
&& \left.+\frac{i}{2}\theta^{mj}\left\{\left(D_{m}\tilde{\psi}\right)^{\dag}\left(D_{i}D_{j}\tilde{\psi}\right) + {\mathrm{c.c}}\right\} \right]
\label{50}
\end{eqnarray}
 Note that $ \hat j_{0}$ does not have the standard form because of the presence of the $\theta$-dependent term. However, it can be brought to almost standard form upto a $\left( 1 - \frac{\theta B}{2}\right)$ factor (assuming to be positive) by dropping a total divergence term, so that it takes a canonical form
\begin{eqnarray}
\int d^{2}x \hat j_{0} = \left(1 - \frac{\theta B}{2} \right)\int d^{2}x \psi^{\dag}\psi = \int d^{2}x \tilde \psi^{\dag}\tilde \psi
\label{51}
\end{eqnarray}
when rewritten in terms of renormalised wave--function $\tilde \psi$ (\ref{4304}). Now since the above expression corresponds to the total charge of a single 
particle, it can be set to unity ($\int d^2 x \tilde \psi^{\dag}\tilde \psi =1$). With this normalisation condition, it now becomes clear that $\tilde{\psi}^{\dag}\tilde{\psi}$ has to be identified as the probability density which is manifestly positive definite at all points.
It immediately follows that the spatial components of $\hat j_{\mu}$, i.e $\hat j_{i}$ {\it must} correspond to the spatial component of the probability current, as $\hat j_{\mu}$ satisfies the continuity equation $\partial_{\mu}\hat j_{\mu}=0$. 
\section{Galilean symmetry generators}
In this section we shall construct all the Galilean symmetry generators for the model defined by the action (\ref{4307}).
 The canonically conjugate momenta corresponding to $\tilde{\psi}$ and $\tilde \psi^{\dag}$ are $\Pi_{\tilde{\psi}}  = \frac{i}{2} \tilde \psi^{\dag}$ and
$\Pi_{\tilde{\psi}^{\dag}}  = -\frac{i}{2} \tilde \psi$.
The  Hamiltonian computed by a Legendre transform reads, 
\begin{eqnarray}
H =&& \int d^{2}x \left[\frac{1}{2\tilde{m}}(D_{i}\tilde{\psi})^{\dag}(D_{i}\tilde{\psi}) - \frac{i}{4}\theta^{mj}(\tilde{\psi^{\dag}}\stackrel{\leftrightarrow}{D_{j}}\tilde{\psi})F_{m0}\right.\nonumber\\ && -\left. A_{0}(\tilde{\psi^{\dag}}\tilde{\psi})\right]
\label{G3}
\end{eqnarray}
 It is clear that the system contains second-class constraints which can be strongly implemented by Dirac scheme \cite{dir} to obtain the following bracket $\left\{\tilde \psi(x), \tilde \psi^{\dag}(y)\right\} = -i \delta^{2}(x - y)$
which in turn can be elevated to obtain the quantum commutator (\ref{4305}).
Now it can be easily checked using (\ref{4305}), the above Hamiltonian (\ref{G3}) generates appropriate time translation
 $\dot {\tilde\psi}(x) = \left\{\tilde\psi(x), H\right\}$.

We can now easily construct the generator of spatial translation and $SO(2)$ rotation by using Noether's theorem and the above mentioned constraints to get, 
\begin{eqnarray}
P_{i} &=&\int d^{2}x\frac{ i}{2} \tilde{\psi}^{\dag}(x)\stackrel{\leftrightarrow}{\partial_{i}}\tilde{\psi}(x)\nonumber\\
J &=& \frac{i}{2}\int d^{2}x \epsilon_{ij}x_{i} \tilde{\psi}^{\dag}(x)\stackrel{\leftrightarrow}{\partial_{j}}\tilde{\psi}(x)
\label{G5}
\end{eqnarray}
which generates appropriate translation and rotation:
\begin{equation}
\left\{\tilde\psi(x), P_{i}\right\}= \partial_{i} {\tilde\psi}(x); 
\left\{\tilde\psi(x),J\right\}=\epsilon_{ij} x_{i}\partial_{j} {\tilde\psi}(x)
\label{G7}
\end{equation}
Note that $J$ consists of only the orbital part of the angular momentum as in our simplistic treatment we have ignored the spin degree of freedom for the field $\tilde\psi$, so that it transforms as an $SO(2)$ scalar. Using the Dirac bracket between $\tilde\psi$ and $\tilde\psi^{\dag}$, one can verify the following algebra:
\begin{eqnarray}
\left\{P_{i},P_{j}\right\} = \left\{P_{i},H\right\} = \left\{J,H\right\} =0 ; \left\{P_{k},J\right\} = \epsilon_{kl}P_{l}
\label{G81}
\end{eqnarray} 
This shows that $P_{k}$ and $J$ form a closed $E(2)$ (Euclidian) algebra. 
Now coming to the boost, we shall try to analyse the system from first principle and shall check the covariance of (\ref{48}) under Galileo boost. For this, we essentially follow \cite{bcasm}. To that end, consider an infinitesimal Galileo boost along the $X$-direction, $t \mapsto t^{\prime} = t,\quad x^{1}{\mapsto} x^{1}{}^{\prime} = x^{1} - vt,\quad x^{2} \mapsto x^{2}{}^{\prime} = x^{2}$,
with an infinitesimal velocity parameter ``$v$''.
The canonical basis corresponding to unprimed and primed frames are thus given as $(\partial / \partial t, \partial / \partial x^{i} )$ and $(\partial / \partial t^{\prime}, \partial / \partial x^{i}{}^{\prime}) $, respectively. They are related as 
\begin{eqnarray}
\frac{\partial }{\partial t^{\prime}} = \frac{\partial }{\partial t} + v \frac{\partial }{\partial x^{1}},\quad \frac{\partial }{\partial x^{i}{}^{\prime}} = \frac{\partial }{\partial x^{i}}
\label{G10}
\end{eqnarray}
Now, note that in the first quantised version $\tilde \psi$ is going to represent probability amplitude and  $\tilde\psi^{\dag}\tilde\psi$ represents the probability density. Hence in order that $\tilde\psi^{\dag}\tilde\psi$ remains invariant under Galileo boost ($\tilde\psi^{\prime \dag}(x^{\prime},t^{\prime})\tilde\psi^{\prime}(x^{\prime},t^{\prime}) =\tilde \psi^{\dag}(x,t)\tilde\psi(x,t)$), we expect $\tilde \psi$ to change atmost by a phase factor. This motivates us to make the following ansatz :
\begin{eqnarray}
\tilde\psi\left(x,t\right) \mapsto \tilde\psi^{\prime}\left(x^{\prime},t^{\prime}\right) = e^{i v \eta(x,t)}\tilde \psi \left(x,t\right) \nonumber\\
\simeq \left(1 + i v \eta \left(x,t\right) \right)\tilde\psi \left(x,t\right)
\label{G11}
\end{eqnarray}
for the transformation of the field $\tilde \psi$ under infinitesimal Galileo boost ($v<< 1$). Further the gauge field $A_{\mu}(x)$ should transform like the basis $\frac{\partial }{\partial x^{\mu}}$ (\ref{G10}).
 This is because $A_{\mu}(x)$'s can be regarded as the components of the one-form $A(x) = A_{\mu}(x)dx^{\mu}$.
 It thus follows that
\begin{eqnarray}
A_{0}(x)\mapsto  A_{0}{}^{\prime}(x^{\prime}) &=&A_{0}(x)+ v A_{1}(x)\nonumber\\A_{i}(x)\mapsto  A_{i}{}^{\prime}(x^{\prime}) &=& A_{i}(x)
\label{G12}
\end{eqnarray}
under Galileo boost.
 Now demanding that
 the equation of motion (\ref{48}) remains covariant implies that the following pair of equations $K\tilde\psi = 0 \quad; K^{'}\tilde\psi^{'} = 0 $
must hold in unprimed and primed frames respectively. Now making use of (\ref{G10},\ref{G11}) in the above equations and then using (\ref{G12}), we get the following condition involving $\eta$ :
\begin{eqnarray}
D_{1}\tilde \psi + i \partial_{0}\eta \tilde \psi =&& \left[-\frac{1}{\tilde m} \partial_{j} \eta - \frac{\theta}{2} \epsilon^{ij} F_{i1}\right] D_{j} \tilde \psi \nonumber\\ && +
\left[- \frac{1}{2\tilde m} \nabla^{2}\eta - \frac{\theta}{2} \epsilon^{ij} F_{i0}\partial_{j} \eta\right]\tilde \psi
\label{G15}
\end{eqnarray}
Since we have considered the boost along the $x$-axis the variable $\eta$ occuring in the phase factor in (\ref{G11}) will not have any $x^{2}$ dependence ($\partial_{2}\eta = 0$). Also since we have taken the background electric field $F_{i0}=E_{i}$ to be constant, we have to consider here two independent possibilities : $\vec{E}$ along the direction of the boost and $\vec{E}$ perpendicular to the direction of the boost. Let us consider the former possibility first. Clearly in this case the term $\epsilon^{ij} E_{i}\partial_{j} \eta $ in the right hand side of (\ref{G15}) vanishes and the above equation becomes 
\begin{eqnarray}
D_{1}\tilde \psi + i \left(\partial_{0}\eta\right)\tilde \psi &=& \left[-\frac{1}{\tilde m} \partial_{1} \eta - \frac{\theta B}{2} \right] D_{1} \tilde \psi \nonumber\\ &&- \frac{1}{2\tilde m}\left(\partial_{1}^{2}\eta\right) \tilde \psi
\label{G16}
\end{eqnarray}
Equating the coefficients of $D_{1}\tilde \psi$ and $\psi$ from both sides we get the following conditions on $\eta$.
\begin{eqnarray}
\left[\frac{1}{\tilde m} \partial_{1} \eta + \frac{\theta B}{2} \right] = -1;\quad i\partial_{0}\eta = - \frac{1}{2\tilde m}\partial_{1}^{2}\eta
\label{G17}
\end{eqnarray}
It is now quite trivial to obtain the following time-independent ($\partial_{0}\eta = 0$) real solution for $\eta$ :
\begin{eqnarray}
\eta = - \tilde m \left(1 + \frac{\theta B}{2}\right) x^{1}
\label{G19}
\end{eqnarray}
This shows that boost in the direction of the electric field is a symmetry for the system. This is, however, not true when electric field is perpendicular to the direction of the boost. This can be easily seen by re-running the above analysis for this case, when one gets 
\begin{eqnarray}
\left[\frac{1}{\tilde m} \partial_{1} \eta + \frac{\theta B}{2} \right] = -1; i\partial_{0}\eta = - \frac{1}{2\tilde m}\partial_{1}^{2}\eta + \frac{\theta E}{2}\partial_{1}\eta
\label{G20}
\end{eqnarray} 
 Clearly this pair does not admit any real solution. In fact, the solution can just be read off as 
\begin{eqnarray}
\eta = - \tilde m \left(1 + \frac{\theta B}{2}\right) x^{1} + \frac{i}{2} \theta E \tilde m t
\label{G21}
\end{eqnarray}
This complex solution of $\eta$ implies the wave function (\ref{G11}) does not preserve its norm under this boost transformation as this transformation is no longer unitary. This demonstrates that the boost in the perpendicular direction of the applied electric field is not a symmetry of the system. Clearly this is a noncommutative effect as it involves the NC parameter $\theta$. This violation of boost symmetry rules out the possibility of Galilean symmetry, let alone any exotic Galilean symmetry obtained by \cite{horvs} in their model.

\section{Hall Conductivity in commutative variables}

In this section, we address the Hall problem \cite{ko} in terms of commutative variables and solve (\ref{48}) in the Landau gauge $A_{0} = E x^{1}, A_{1} = 0, A_{2} = B x^{1}$. Taking the trial solution $\tilde \psi(t, x^{1}, x^{2}) = e^{-i\omega t}e^{ip_{2} x^{2}}\phi(x^{1})$, we obtain, after appropriate change of variables, the standard HO equation with an enhanced frequency $\tilde \omega_{c} = (1+ \theta B)\omega_{c}$,
\begin{eqnarray}
\left[- \frac{1}{2\tilde{m}}\partial_{\bar X}^{2} + \frac{\tilde{m} \tilde{\omega}_{c}^{2}}{2} \bar {X} ^{2} \right]\tilde\phi\left(\bar X\right) = \xi \tilde\phi\left(\bar X\right)  
\label{550}
\end{eqnarray}
where, $\tilde\phi(\bar{X}) = \phi(x^{1})$, $\bar{X}  = (x^{1} - \frac{p_{2} + \tilde{m}E/B}{B} -\frac{\tilde{m}E \theta}{2B})$ and $\xi= \omega + p_{2}E/B +\frac{\tilde{m}}{2}\left(E/B\right)^{2} + \frac{\tilde{m}}{2}\theta\left(E^{2}/B\right) $, is the harmonic oscillator energy eigen-value. The eigen-functions are given in terms of Hermite polynomials with the admissible values of $\xi$ given by $\xi_{n} = (n + \frac{1}{2})\tilde \omega_{c}$. This implies a quantisation condition for $\omega$
as $\omega_{n}=(n + \frac{1}{2})\tilde \omega_{c} -\left[ \left( p_{2}E/B + \frac{\tilde{m}}{2}\left(E/B\right)^{2} + \frac{\tilde{m}}{2}\theta\left(E^{2}/B\right)\right)\right]$. Interestingly, the above spectrum changes drastically under $B\mapsto-B$ showing parity violation. This feature is also there in the commutative case ($\theta=0$). However, under $x^{1} \mapsto -x^{1}$, $x^{2} \mapsto x^{2}$, there is no change in the spectrum, as both $E$ and $\theta$ flip sign along with the $B$.
Now since $\hat j_{1} = 0$, corresponding to the above wave-function, the longitudinal current vanishes and the transverse current for a single particle $I^{(1)}_{2}=\int dx^{1} \hat j_{2}$ can be written compactly as,
\begin{eqnarray}
I^{(1)}_{2} &=& \int dx^{1} \frac{1}{2\tilde{m}i}\left(1 - \frac{\theta B}{2}\right)\left[\tilde{\psi}^{\dag}\left(D_{2}\tilde{\psi}\right) -\left(D_{2}\tilde{\psi}\right)^{\dag}\tilde{\psi} \right]\nonumber\\
&=&-\int d\bar{X} \frac{E}{B} |\tilde\phi(\bar{X})|^{2} = - \frac{1}{L_{y}}\left(\frac{E}{B}\right)
\label{582}
\end{eqnarray}
where we have used $\tilde \psi^{\dag}D_{2}\tilde \psi = -(D_{2}\tilde \psi)^{\dag}\tilde \psi = i(p_{2} - A_{2})(\phi(x^{1}))^{2}$ and
 the normalisation condition $\int d^2 x \tilde\psi^{\dag}\tilde\psi=\int d\bar{X}dx^{2} |\tilde\phi(\bar{X})|^{2}=1$ which for a sample width
$L_{y}$ yields the condition $\int d\bar{X}|\tilde\phi(\bar{X})|^{2} = 1/L_{y}$.
 Now following \cite{st}, we multiply $I^{(1)}_2$ by the number of available states $\rho L_{x}L_{y}$ in a rectangular area $L_{x}L_{y}$, ($\rho$ is the density of states), to get the total current as $I=-\frac{\rho}{B}V$, ($V=E L_{x}$ is the potential drop along the $x$-axis). This yields the standard Hall conductivity expression (involving the filling fraction $\nu$), $\sigma_{H}=-\rho/B =-\nu/2\pi$, without any $\theta$-correction.

\section{Conclusions}

In this paper we have obtained an effective $U(1)$ gauge invariant Schr\"{o}dinger action by using SW map followed by wave-function and mass renormalisation. The effect of non-commutativity on the mass parameter appears naturally in our analysis. Interestingly, we observe that the external magnetic field has to be static and uniform in order to get a canonical form of Schr\"{o}dinger equation upto $\theta$-corrected terms, so that a natural probabilistic interpretation emerges. The Galilean symmetry of the model is next investigated where the translation and the rotation generators are seen to form a closed Euclidean sub algebra of Galilean algebra. However, the boost is not found to be a symmetry of the system, even though the condition $\theta^{0i} = 0$ is Galilean invariant. Finally, we compute Hall conductivity which turned out to have no $\theta$-correction.


\section*{Acknowledgment}

We would like to thank R.Banerjee and J.K.Bhattacharjee for useful discussions. AS would like to thank the Council for Scientific and Industrial Research (CSIR), Govt. of India, for financial support. We also thank the referee for useful comments.

\end{document}